\documentclass[preprint]{aastex63}

\usepackage{xcolor}
\usepackage{physics}

\newcommand{\ACen}{$\upalpha$ Cen\ }
\newcommand{\ACene}{$\upalpha$ Cen}

\renewcommand\arcsec{\mbox{$^{\prime\prime}$}}%

\newcommand{\PSUAA}{Department of Astronomy \& Astrophysics, Pennsylvania State University, University Park, PA, 16802, USA}
\newcommand{\PSUCEHW}{Center for Exoplanets and Habitable Worlds, Pennsylvania State University, University Park, PA, 16802, USA}
\newcommand{\PSETI}{Penn State Extraterrestrial Intelligence Center, Pennsylvania State University, University Park, PA 16802, USA}
\newcommand{\PSARC}{Penn State Astrobiology Research Center, Pennsylvania State University, University Park, PA 16802, USA}
\newcommand{\PSUCHEM}{Department of Chemistry, Pennsylvania State University, University Park, PA 16802, USA}
\newcommand{\PSUGEO}{Department of Geosciences, Pennsylvania State University, University Park, PA 16802, USA}
\newcommand{\UDELPA}{Department of Physics and Astronomy, University of Delaware, Newark, DE 19716, USA}
\newcommand{\UCB}{Department of Astronomy, University of California, Berkeley, Berkeley, CA, 94720, USA}
\newcommand{\SETIi}{SETI Institute, Mountain View, California}
\newcommand{\UCI}{Department of Physics \& Astronomy, The University of California, Irvine, Irvine, CA 92697, USA}

\usepackage{amsmath}
\usepackage{amsfonts}
\usepackage{amssymb}
\usepackage{upgreek}
\usepackage{graphics}
\usepackage{graphicx}
\usepackage{float}
\usepackage{changepage}
\usepackage{framed}
\usepackage{}

\usepackage{natbib}
\graphicspath{{./}{figures/}}

\begin{document}

\title{A Search for Radio Technosignatures at the Solar Gravitational Lens Targeting Alpha Centauri}


\author[0000-0001-9686-5890]{Nick Tusay}
\affiliation{\PSUAA}
\affiliation{\PSUCEHW}
\affiliation{\PSETI}

\author[0000-0003-4591-3201]{Macy J. Huston} 
\affiliation{\PSUAA}
\affiliation{\PSUCEHW}
\affiliation{\PSETI}

\author[0000-0001-9408-8848]{Cayla M. Dedrick} 
\affiliation{\PSUAA}
\affiliation{\PSUCEHW}

\author[0000-0003-2633-2196]{Stephen Kerby} 
\affiliation{\PSUAA}

\author[0000-0002-4677-8796]{Michael L. Palumbo III}  
\affiliation{\PSUAA}
\affiliation{\PSUCEHW}

\author[0000-0003-4823-129X]{Steve Croft}
\affiliation{\UCB}
\affiliation{\SETIi}

\author[0000-0001-6160-5888]{Jason T. Wright}
\affiliation{\PSUAA}
\affiliation{\PSUCEHW}
\affiliation{\PSETI}

\author[0000-0003-0149-9678]{Paul Robertson}
\affiliation{\UCI}

\author[0000-0001-7057-4999]{Sofia Sheikh}
\affiliation{\SETIi}
\affiliation{\PSETI}

\nocollaboration{200}


\author{Laura Duffy}
\affiliation{\PSUAA}

\author[0000-0002-2944-6060]{Gregory Foote} 
\affiliation{\UDELPA}

\author{Andrew Hyde}
\affiliation{\PSUGEO}

\author{Julia Lafond}
\affiliation{\PSUGEO}

\author{Ella Mullikin} 
\affiliation{\PSUCHEM}

\author[0000-0001-7142-2997]{Winter Parts}
\affiliation{\PSUAA}
\affiliation{\PSUCEHW}

\author[0000-0002-6019-4818]{Phoebe Sandhaus} 
\affiliation{\PSUAA}
\affiliation{\PSUCEHW}

\author{Hillary H. Smith}
\affiliation{\PSUGEO}
\affiliation{\PSARC}

\author[0000-0001-5290-1001]{Evan L. Sneed} 
\affiliation{\PSETI}

\collaboration{200}{Penn State Astro 576 SETI Class, Fall 2020}


\author[0000-0002-8071-6011]{Daniel Czech}
\affiliation{\UCB}

\author[0000-0002-8604-106X]{Vishal Gajjar}
\affiliation{\UCB}

\collaboration{200}{Breakthrough Listen}

\correspondingauthor{Macy Huston}
\email{mhuston@psu.edu}

\begin{abstract}
Stars provide an enormous gain for interstellar communications at their gravitational focus, perhaps as part of an interstellar network. If the Sun is part of such a network, there should be probes at the gravitational foci of nearby stars. If there are probes within the solar system connected to such a network, we might detect them by intercepting transmissions from relays at these foci. Here, we demonstrate a search across a wide bandwidth for interstellar communication relays beyond the Sun's innermost gravitational focus at $550 \:\rm{AU}$ using the Green Bank Telescope (GBT) and Breakthrough Listen (BL) backend. As a first target, we searched for a relay at the focus of the Alpha Centauri AB system while correcting for the parallax due to Earth's orbit around the Sun. We searched for radio signals directed at the inner solar system from such a source in the L and S bands. Our analysis, utilizing the {\tt turboSETI} software developed by BL, did not detect any signal indicative of a non-human-made artificial origin. Further analysis excluded false negatives and signals from the nearby target HD 13908. Assuming a conservative gain of $10^3$ in L-band and roughly 4 times that in S-band, a $\sim$1 meter directed transmitter would be detectable by our search above 7 W at 550 AU or 23 W at 1000 AU in L-band, and above 2 W at 550 AU or 7 W at 1000 AU in S-band. Finally, we discuss the application of this method to other frequencies and targets. 
\end{abstract}


\section[Introduction]{Introduction} \label{sec:intro}

The Search for Extraterrestrial Intelligence (SETI) often focuses on looking for signs of technology around other stars. This, however, is just one way to approach the search for ETI. 
\cite{bracewellCommunicationsSuperiorGalactic1960} posited that ``superior communities'' would send material probes to neighboring star systems for the purpose of communication. If this is the case, we should also be searching our own solar system for evidence of ETI. This idea is the basis of artifact SETI, or solar system SETI, \citep[also called SETA, the Search for Extraterrestrial Artifacts;][]{seta}, the search for any material artifacts that may be sent by ETI during a long-term galactic exploration effort. \citet{seta} proposed that a search for such objects in our solar system could be carried out using telescopes, radar, infrared radiation, direct probes, and other methods. 

\subsection[The Interstellar Communication Network Hypothesis]{The Interstellar Communication Network Hypothesis}

Communications between probes and their home systems across interstellar distances would require extraordinarily high directional gains or transmitting powers to ensure that a distant receiving dish can reliably reconstruct information packets. \cite{eshlemanGravitationalLensSun1979} proposed that the Sun's gravitation could be used as a lens to magnify the radiation to or from a distant source. A spacecraft beyond 550 AU along the focal line could use this effect to observe and communicate at interstellar distances using instruments comparable to those that we currently use for interplanetary communication. One proposed application of this mechanism is to utilize the Solar Gravitational Lens (SGL) to provide huge magnification to a telescope along the Sun's focal line at a distance between 550 and 1000 AU \citep[e.g. the FOCAL mission; ][]{Maccone1994,maccone2010,Maccone1997}.

\cite{Gillon2014} proposed a new SETI approach to monitor the focal regions of the SGL corresponding to nearby stars to search for communicative technology. Their hypothesis is based on two assertions about galactic exploration. First, ``Von Neumann'' (self-replicating) probes \citep{Tipler1980} exploring the Galaxy would require some communication among probes and back to the original system. Second, in order to communicate across these interstellar distances, the explorers would create a network of probes leveraging stars as gravitational lenses. These assumptions solve the problem of long-distance communication across the Galaxy by eliminating long-range communication in favor of a locally multi-nodal network connected to their nearest neighbors \citep{Gertz2017,Gertz2020}; this strategy is commonly used for communication networks around the Earth, such as the Internet.
\cite{Gillon2014} acknowledged that such devices would not likely be detectable through imaging or stellar occultation, but that multi-wavelength monitoring may detect leaked signals. 

\cite{Hippke2020a,Hippke2020b,Hippke2021} explored technical constraints involved in the configuration of an interstellar communication network using stellar gravitational lenses, in order to motivate searches at the proper locations and sizes for nodes and probes. Their work argued that optimal probe sizes would be on the order of 1 meter, the optimal communication wavelength range would be from 100 $\mu$m to 1 nm, and that a distance of 1,000 AU from the Sun is preferred. \citet{Hippke2020b} also notes that, in this proposed configuration, separate exploration probes would be required in order to study the inner solar system. \cite{Kerby2021} explore the engineering requirements and sustainability for an SGL relay to remain in proper position, finding that such technology is feasible and noting that another observable property of such probes may be the by-products of station-keeping propulsion. They also argued that single stars (i.e.\ those with minimal gravitational perturbations from companions) were the best hosts for such probes because their station-keeping costs were smallest.

\cite{Gillon2021} searched for a probe on the Solar focal line in communication with the Wolf 359 system, the third nearest star to our Sun and one from which Earth's transit across the Sun would be observable, at a time when Earth could have intercepted the transmissions. Their search for optical signals from the solar system toward the star, as well as for an object with the position and motion hypothesized within the extent of Uranus' orbit (20 AU), did not reliably identify any such probe. 

\subsection[Solar Gravitational Lensing for Interstellar Communication]{Solar Gravitational Lensing for Interstellar Communication}\label{sec:Gain}

\begin{figure}
    \centering
    \includegraphics{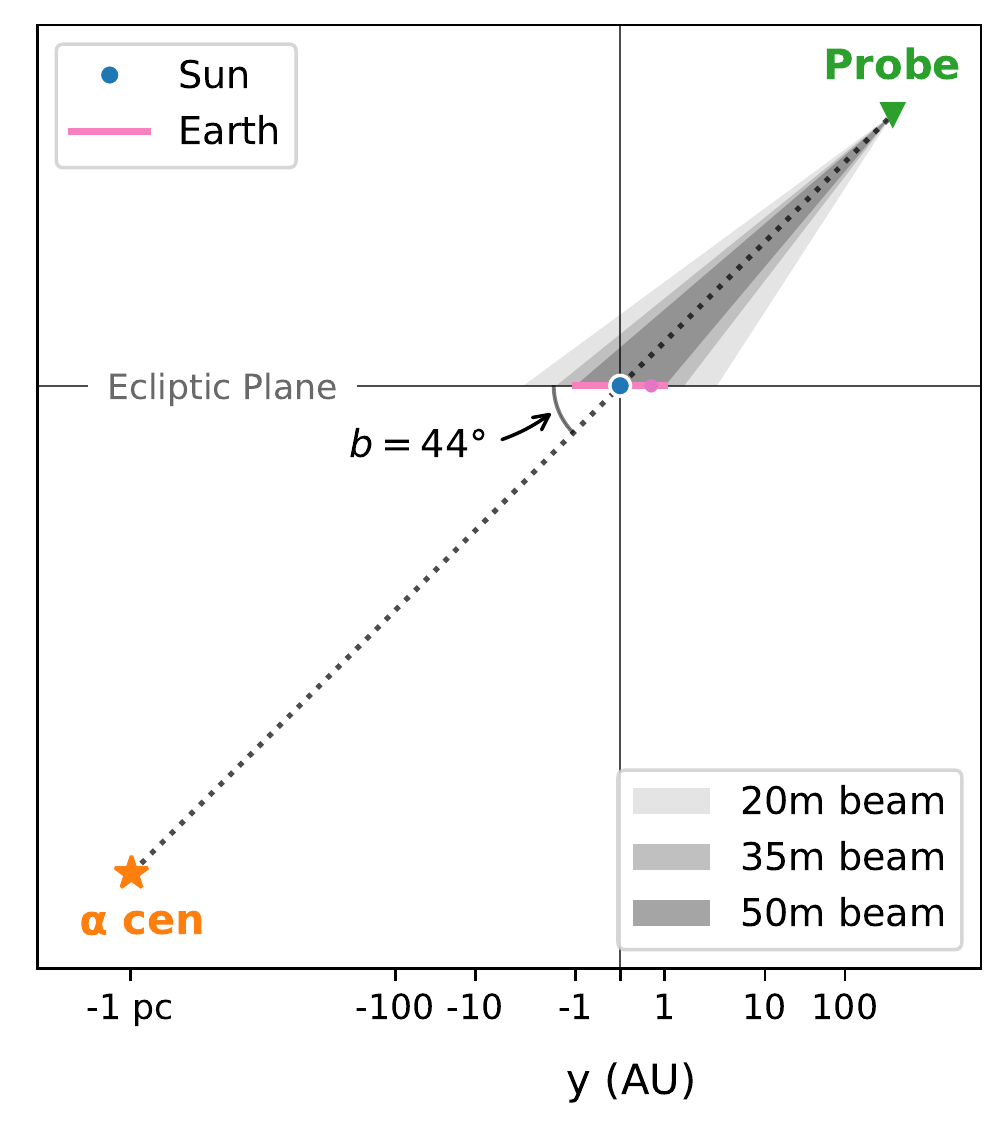}
    \caption{Visualization of an SGL probe opposite \ACene. The Sun is at the origin in our coordinate system, and Earth's orbit lies on the xy\-plane. The probe is indicated in green and its outgoing beams for various dish sizes in gray (see Section \ref{sec:Gain}).}
    \label{fig:setup}
\end{figure}

Massive objects in the universe, such as black holes and stars, bend the trajectories of nearby photons. This process can create a lensing effect similar to a focusing element of a telescope \citep{einsteinLensLikeActionStar1936}.
Gravitational lensing warps a source object's apparent shape into two distorted images. The distortion effect centers on a ring around the lens object, called the Einstein ring, which has an angular radius of:
\begin{equation}
    \theta_E = \left( \frac{4GM D_{\rm LS}}{c^2 D_{\rm L} D_{\rm S}} \right)^{1/2},
\end{equation}
where $G$ is the gravitational constant, $M$ is the mass of the lens object, $c$ is the speed of light, $D_{\rm L}$ is the observer-lens distance, $D_{\rm S}$ is the observer-source distance, and $D_{\rm LS}$ is the lens-source distance. In a solar gravitational lens system, the Sun is the lens, the distant star is the source, and the probe is our origin/observer. The physical Einstein ring radius ($R_E = D_L \theta_E$) of the solar lens should be at least the radius of the sun. This works out to a minimum $D_L$ of roughly 550 AU, which we adopt as our minimum possible probe distance. Figure \ref{fig:setup} shows the layout for a probe utilizing the SGL to send or receive messages to/from \ACen. We note that \cite{Gillon2021} argue that the use of out-of-focus nodes at lower separations may also be worthwhile.

It may be possible to detect signals from a relay spacecraft at a distance $d \gtrsim 550 \:\rm{AU}$ from the Sun if the Earth is contained within the opening angle of the probe's outgoing transmission beam at any point in its orbit. For target stars that lie along the ecliptic plane, the Earth will always pass through the beam when it transits and eclipses the Sun. However, from the view of a relay off of the ecliptic plane, Earth will always have some non-zero separation from the Sun, which its beam may not encapsulate. From the point of view of a probe at a distance of 550 AU, the maximum angular separation between the Earth and the Sun is 6.3\arcmin, regardless of its ecliptic latitude. A probe with an ecliptic latitude of $b$ sees a minimum angular separation 
\begin{equation} \label{eq:incl}
    \phi_{\rm E, min} \approx \left( \frac{d}{550~\rm{AU}} \right)^{-1} \sin(b) \times 6.3~\rm{arcmin} .
\end{equation}

The Earth's entire orbit would be in a beam for 10-cm wavelength signals for any transmitter with diameter $\lesssim$ 34 m. However, Earth's point of closest approach to the SGL line may fit into smaller beams. This is a unique feature of performing this search search at radio wavelengths. In optical wavelengths, the focus of prior searches, the beam sizes for dishes on the order of 10m would be much less than 6.3\arcmin. Thus, eavesdropping on optical SGL signals would only be feasible for target stars in the ecliplic plane. Following Equation \ref{eq:incl}, for a probe opposite \ACen ($b\sim 44^\circ$), the minimum visible beam size at the point of closest approach would be 4.4\arcmin. This corresponds to transmitter diameter $\lesssim$ 50 m. 

The directional gain for a transmitting relay is
\begin{equation}
    G_{\rm SC} = \frac{4 \pi^2 r^2_{\rm t} k_{\rm t}}{\lambda^2},
\end{equation}
where $r_{\rm t}$ is the radius of the transmitter, $k_{\rm t}$ is its efficiency (which we assume to be of order unity), and $\lambda$ is the emitting wavelength \citep{macconeInterstellarRadioLinks2011}. 
\citet{Hippke2020b} suggests $\sim$1 meter transmitter size for sub-mm signals, but a larger transmitter would be required to produce similar gains at radio wavelengths. A 50m transmitter could achieve directional gains on the order of 65 dB, and even a 20 m transmitter could achieve 55 dB.

From Equations 8-9 of \cite{macconeInterstellarRadioLinks2011}, we find that the gain resulting from a solar lens is:
\begin{equation}
    G_\odot = \frac{8 \pi^2 G M_\odot}{c^2 \lambda} .
\end{equation}
For $\lambda \sim$ 10 cm, the solar lens gain is roughly 60 dB. Combining the directional gain of the relay probe and the lensing of the Sun, a transmitting spacecraft could achieve a total gain of $G_{\rm tot} = G_\odot \times G_{\rm SC}$ of over $120\:\rm{dB}$, overcoming the difficulties of transmission across interstellar distances by focusing transmissions into tight parallel beams.

\subsection[Searching for Technology at the SGL]{Searching for Technology at the SGL}

\subsubsection[What to Look for]{What to Look for} \label{sec:whatlookfor}

There have been many suggestions in the literature of purposes for probes in the solar system. The SGL foci of nearby stars are special locations for probes because they allow access to a ``Galactic Internet," but the purpose and functions of such probes is otherwise unconstrained. In contrast to classical radio SETI searches for intentional messages sent toward Earth from interstellar distances, this method primarily attempts to eavesdrop on communications between two technological structures. We offer here three suggestions from the literature on such functions and the resulting implications for our search, but emphasize that our search is not dependent on and does not assume the probes we seek serve such functions.

Direct interception of outgoing communication through the SGL is the obvious signal to search for under the interstellar communication network hypothesis, but it suffers from some technical limitations. For interstellar transmissions from a relay spacecraft to be detectable from the Earth, the stellar relay must be actively transmitting when we search, such searches can only be performed when the Earth happens to be in the beam, and/or the beam could be tightly focused on a portion of the Einstein ring the Earth does not transit. A constantly transmitting stellar relay is more likely to be detected from Earth than a relay that only intermittently sends signals to the target star or local probes. 

When searching for a relay, one major obstacle is determining which wavelengths would be most appropriate for SGL lensing. Radio wavelengths are an obvious choice for interstellar communication, due to their low energy and extinction. However, the solar corona can interfere with long wavelength radio signals. \cite{Turyshev2019} found that photons with $\lambda \geq$ 157 cm ($\nu \leq$ 0.2 GHz) are completely blocked. At shorter wavelengths, plasma refraction steers some photons away from the focal line, decreasing signal strength \citep{Hippke2020b}. Moving from 3 cm to 200 $\mu$m (10 GHz to 1.5 THz), the effect gradually decreases. At wavelengths below $\sim 100 \mu$m ($\nu \gtrsim$ 3 THz), the decrease in solar lens gain is negligible. For wavelengths non-negligibly but not fully blocked, the SGL could still provide significant gains, so these are not completely eliminated from consideration, but do make searches for intercepted transmissions at radio or microwave frequencies less well motivated. 

On the other hand, shorter wavelengths produce smaller beam widths, which would make eavesdropping on an SGL signal impossible for reasonable transmitter sizes and targets not near the ecliptic plane.

\begin{figure}[!ht]
    \centering
    \includegraphics[width=0.7\textwidth]{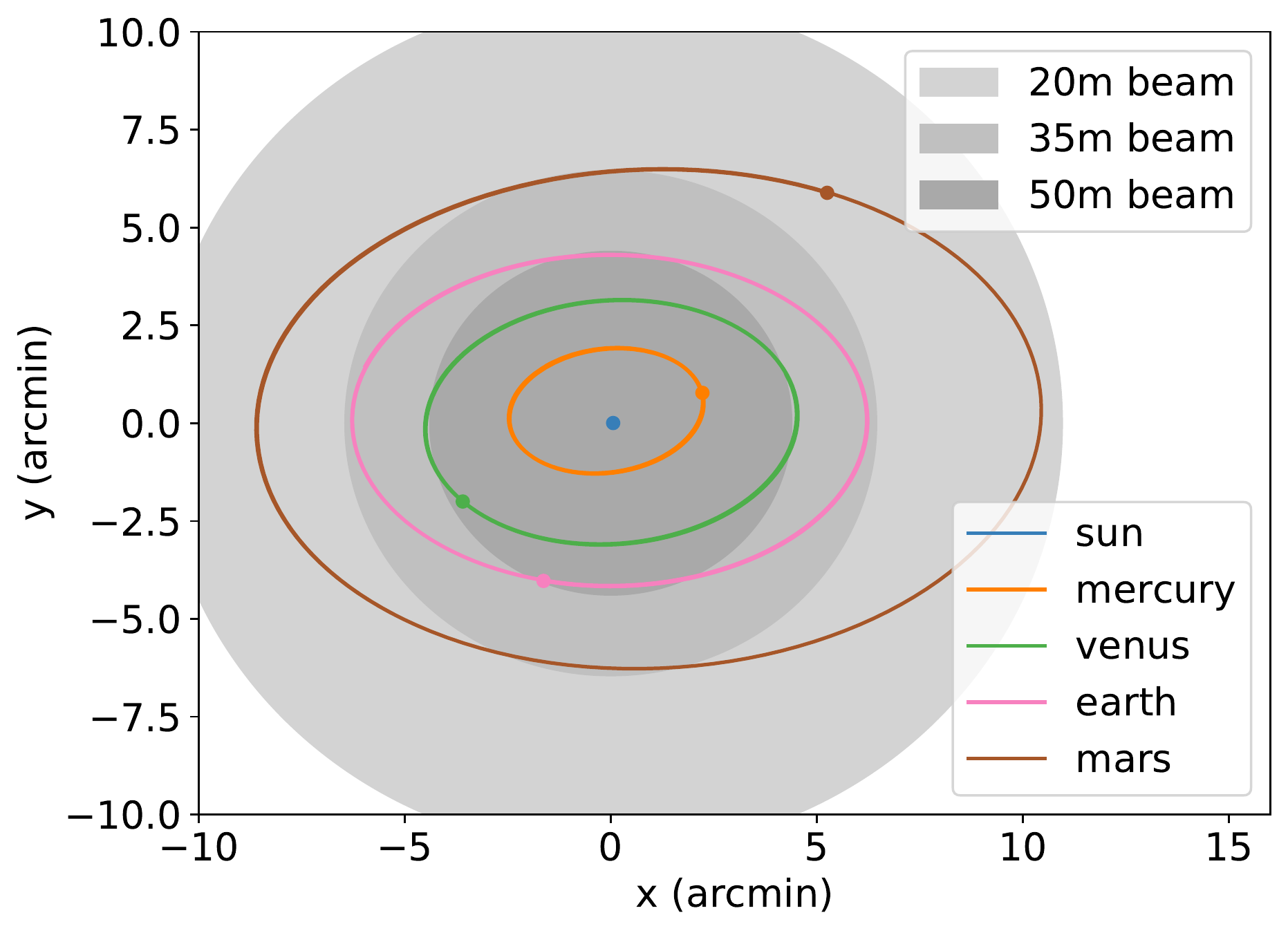}
    \caption{View of the solar system from a probe opposite \ACen at 550 AU on the night of our observations. The Sun lies at the origin, and the inner solar system planets' orbits are shown. Gray filled circles indicate beam widths for various transmitter diameters at 10-cm wavelengths. For a probe located at a distance $d~>$ 550 AU, these orbital axes would scale down by a factor of $d$/(550 AU).}
    \label{fig:solarsystem}
\end{figure}

In addition to sending messages via the SGL, a relay spacecraft might send signals to other probes in the inner solar system. Indeed, such a scheme is recommended by \cite{Maccone2022} as a way to retrieve information from interstellar probes, and is reminiscent of relay schemes used by humans' interplanetary probes, for instance for communication with Mars landers. There is no reason such secondary communications might not happen at different wavelengths than the interstellar communications through the SGL, since coronal interference is not a problem for them. It is thus possible they would occur at radio frequencies and with beam sizes and directions that would allow them to be intercepted at Earth year-round. Figure \ref{fig:solarsystem} shows the view of the solar system from a probe at the Sun's focal point for communication with \ACene. 

\cite{bracewellCommunicationsSuperiorGalactic1960} suggested that the purpose of a solar system probe would be to serve as a beacon, and \cite{seta} and others considered places where such beacons would reside. Success in intentional communication SETI requires a determination of how to find another group of beings without being able to communicate beforehand. This requires the determination of optimal places, times, frequencies, and other aspects of possible communication which may be chosen by the other group. In game theory, these optimal choices are referred to as ``focal points'' \citep{Schelling60}, but we use the term ``Schelling points'' in order to avoid confusion with the optical definition of focal points \citep{Wright2020}. The SGL focal line is a recognizable location for both humans and the residents of nearby stars, and it is a relatively static location in the sky. Thus, it may be an optimal place to place a beacon to intentionally attempt to make contact with technological life in the Solar System.  

Beacons sent to other star systems may be a preferred method of interstellar first contact, as it does not immediately reveal the location of the sender, reducing risk of aggressive retaliation \citep{Gertz2018}. An interstellar communication relay only reveals the position of the next node in the web. By their nature, beacons are intended to be found, so a beacon at the SGL might transmit at any frequency its constructors thought potential recipients would guess, further justifying our choice of frequencies near the ``water hole'' \citep{Oliver1979} for this study.

For this work, our first observations of the SGL for nearby stars as a proof-of-concept, we have chosen to observe in L and S bands. We chose these bands for three reasons: (1) the size of beams at these frequencies allowed us to check them all in a single 1 hour session, (2) these are in or near the ``water hole'' (to search for beacons), and (3) these are or are near frequencies humans use for interplanetary communication (to search for communications with probes in the solar system). 

\subsubsection[Where to Look]{Where to Look}
\label{wheretolook}

To first order, we expect this type of probe to lie at the antipode of some nearby star, around which another interstellar communication network probe could reside. Initially, we neglect light travel time considerations and place the probe at the exact antipode of the target star as we currently see it in the sky. This position on the sky, viewed from Earth, will vary slightly based on the probe's assumed distance from the Sun because, being a small fraction of a parsec distant, the probes suffer significant parallax. 

For observation time $t$, we consider the positions of the Sun and Earth in the barycentric frame, $S(t)$ and $E(t)$ respectively. We also define unit vector $x(t)$ as the direction of the target star from the Sun. We consider Sun-probe distances, $z$, from 550 AU to infinity. We then calculate the barycentric frame position of the probe as:
\begin{equation}
    P = S(t) - z \times x(t) .
\end{equation}
Finally, on-sky coordinates {of the probe} are calculated from the Earth-probe vector $\overrightarrow{PE} = P-E(t)$.

{Finite} light travel time requires some corrections to this expected position \citep{Seto2020}. We must consider the positions of our celestial objects at other points in time. {This calculation can be performed for two probe types: receivers and transmitters.}

{For a probe transmitting signals through the SGL to a star, the signal must travel to the position the star will be one sun-star light travel time in the future.  Since our naive calculation uses the apparent position of the star at our observation time, which is where the star was one travel time ago, we must advance the star's position by two light travel times.}

{In principle, we must also consider that the probe must aim to where the sun will be when the signal arrives, and advance its position by one Sun-probe light travel time.  We neglect this effect because the maximum change it has on the position of the probe is of order 0\farcs 1.}

{We thus calculate the transmitting probe position as:
\begin{equation}
    P(\mathrm{transmitter}) = S(t) - z \times x(t + 2d/c) .
\end{equation}
\noindent where $c$ is the speed of light.
}

{For a probe near the sun receiving signals from a star, our naive calculation of the star's position is appropriate, because the signal will arrive from the apparent (retarded) position of the star, just as other light from the star does. But for a distant probe, we need to consider that we see the probe where it was one Earth-probe light travel time ago, when it was sitting in a position corresponding to the star's position one probe-Sun travel time before that.  Formally, we should then retard the star's apparent position by a time $(z+|PE|)/c$, however for computational simplicity we approximate $|PE|\approx z$ because the difference in times is small compared to the timescale over which the star's apparent position changes significantly.}

{Our receiving probe position is then:
\begin{equation}
    P(\mathrm{receiver}) = S(t) - z \times x(t - 2z/c),
\end{equation}
}

{This light travel time complication means that the probe position does not converge at infinite distance. Here, we assume that such a probe would not be placed further than a tenth of the Sun-star distance.}

For both types of probe we have a locus across the sky of possible locations for a probe at varying distance $z$ from the Sun. In general, the full extent of this line across the sky should be observed in order to detect or rule out active communication from a probe at the target star's SGL location.

\subsection[The SGL for Alpha Centauri]{The SGL for \ACene}

In this work, we demonstrate the SGL relay SETI search method described above with a search for a hypothetical probe in contact with $\alpha$ Centauri. {Figure~\ref{fig:obs} shows the positions of transmitting and receiving probes for \ACen A (the two components of \ACen are separated by less than 10\arcsec, which is so much smaller than our beam that our search encompasses both components). }

The observations are described in Section \ref{sec:acenobs}. We present the analysis of our data in Section \ref{sec:analysis}. In Section \ref{sec:disc}, we discuss the results of our analysis and additional artifact SETI searches of this type. We conclude in Section \ref{sec:concl}. 

\section[Alpha Centauri Observations with GBT]{$\alpha$ Centauri Observations with GBT}
\label{sec:acenobs}

\begin{figure}
\centering
\includegraphics[width=0.8\textwidth]{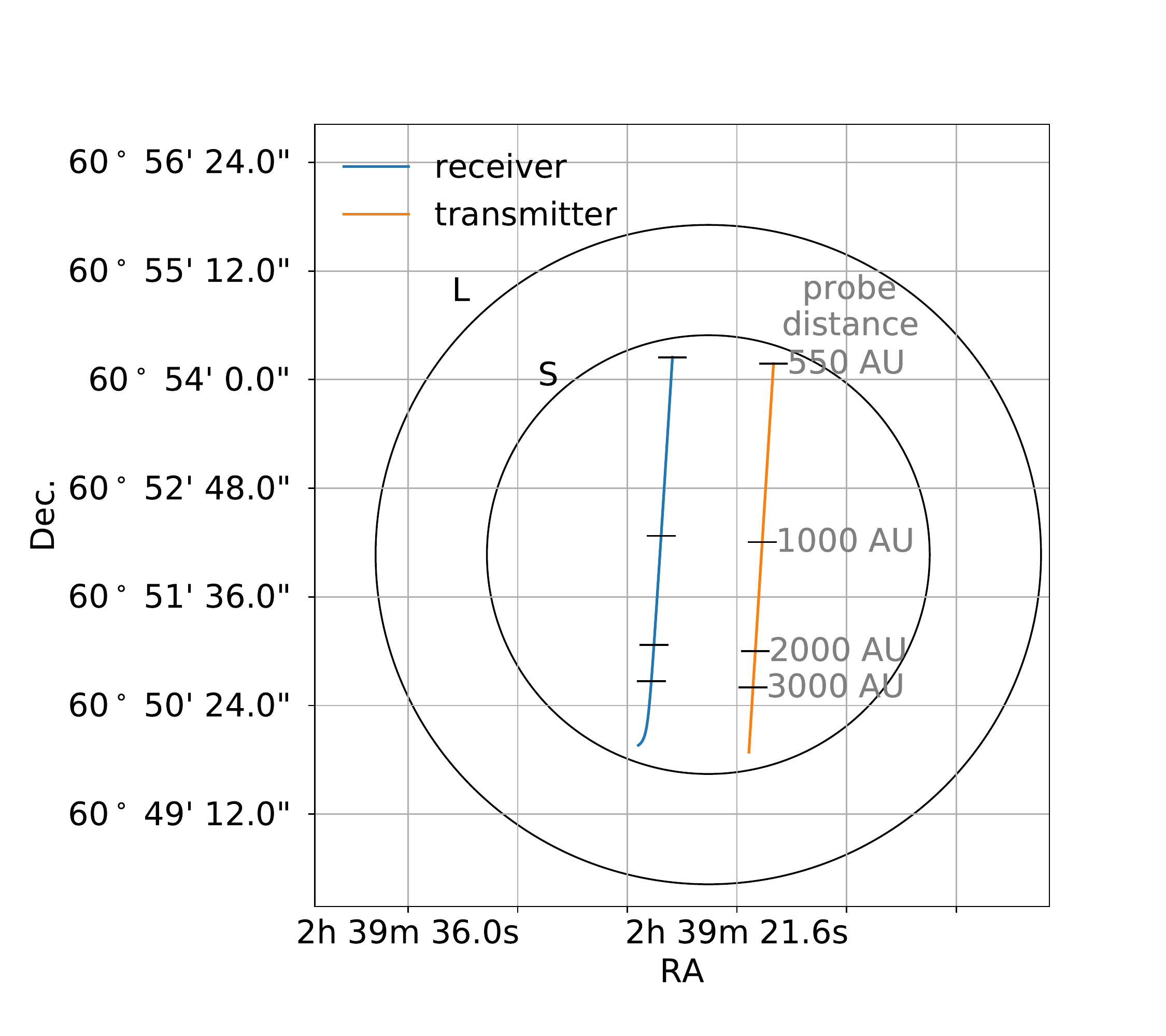}
\caption{Observations of the possible probe positions for communication with \ACen on UT 2021 November 6. The blue and orange lines mark the positions for receiving and transmitting probes, respectively, with light travel time taken into account. Tics along the probe position lines show where a probe at certain distances from the Sun would lie. The FWHM beam positions of each observation are represented by black circles, labelled by filter.}
\label{fig:obs}
\end{figure}

We selected a set of positions corresponding to the possible locations of a relay probe in communication with $\alpha$ Centauri, considering both the receiver position and the transmitter position. We took observations with the Green Bank Telescope (GBT) in the L and S bands using the Breakthrough Listen (BL) backend \citep{MacMahon2018}. As shown in Figure \ref{fig:obs}, we observed the region of the sky directly opposite \ACene, placing observations along a line to account for the parallax of a probe at finite distances from the Earth and Sun. A probe at infinite distance from the Sun with no accounting for light travel time will be positioned exactly opposite \ACen on the sky and have no parallax, but a probe at $550 \:\rm{AU}$ will be offset from the exact antipode by up to six arcminutes, depending on the position of Earth in its orbit. We trace two of these lines, as receiver and transmitter probes require different considerations of the light travel time between the Sun and \ACene, as described in Section \ref{wheretolook}. The GBT pointings were selected to cover regions along the lines of possible positions on the sky, leveraging the different beam width of the various bands in each case.

We observed our target area on UT 2021 November 6. The Earth is at its closest position to the vector connecting the Sun to the target point in early November. In addition to minimizing the beam size required for the hypothesized signal to be detectable from Earth, this also minimizes the number of pointings required to cover the entire line of possible probe locations. 

Data were taken at 300s per scan using an ABABAB pattern. The ABABAB sequence consists of three ``nods" between our ``ON" source target and an ``OFF" target. The OFF target is used to identify and correct for radio frequency interference (RFI). We chose HD 13908 as our OFF target because it has the smallest on-sky separation from the ON target of any known planet-hosting star. The separation between the ON and OFF targets is 317\arcmin, which is equivalent to 37 beam widths at L-band and 58 at S-band. 
Additionally, a scan of a pulsar was completed at the beginning of each night of observations to confirm that the telescope and instrument were performing as expected.

\section[Analysis]{Analysis}
\label{sec:analysis}

\subsection[TurboSETI]{TurboSETI} \label{sec:tseti}
We used {\tt turboSETI} \citep{enriquezTurboSETIPythonbasedSETI2019} to analyze our observations by searching for signals with specific drift rates. In the rest frame of the Sun, the proposed communication relay is effectively stationary, so a monochromatic transmission viewed from Earth should only show Doppler shifts due to the orbital motion and the rotation of the Earth. To obtain a drift rate for a signal from the relay, we used {\tt barycorrpy} \citep{kanodiaBarycorrpyBarycentricVelocity2018} to calculate the radial acceleration between GBT and the probe. Given this radial acceleration $dv_r/dt$, we calculate a maximum drift rate $\dot{f}_{\rm max}$ in ${\rm Hz/s}$:
\begin{equation}
    \dot{f}_{\rm max} = \frac{dv_r}{dt} \frac{f_{\rm max}}{c},
\end{equation}
\noindent where $f_{\rm max}$ is the maximum rest-frame frequency observed.

The magnitude of the drift rate depends on the frequency and relative acceleration at the time of observation. The largest contribution to the relative acceleration term is due to the rotation of the Earth. A signal from a probe at rest relative to the Sun should exhibit a negative drift rate to an observer on Earth \citep{Sheikh:2019:}.

We calculate drift rates for the hypothetical probe relative to the barycentric frame at the time of observation based on the pointing of each beam. These drift rates are around $-0.124$ to $-0.127$ {\rm Hz/s} in L-band, and $-0.18$ {\rm Hz/s} in S-band. To account for uncertainty in the probe's position within a generous margin, we multiply the calculated drift rate for each pointing by a small numerical factor ($\times 1.3$) and use these modified drift rates as the maximum drift rates in our search through the data. Without specifying a minimum, we employ {\tt turboSETI} to search for hits up to this maximum drift rate. 

Since the maximum drift rate of the hypothetical probe is quite small, it is computationally feasible to conduct a high-sensitivity search. By default, {\tt turboSETI} searches for signals above a SNR threshold of 25. For our analysis, we lower this threshold to 10. 

Following an initial search for signals, {\tt turboSETI} performs an additional check on the preliminary detections using three different ``filter thresholds" to remove RFI from the pool of candidates based on the ON/OFF observing cadence. The first filter threshold simply looks for signals above the specified SNR. The second threshold looks for signals above the SNR that appear in at least one on-pointing, but not in any off-pointings. And the third looks for signals that appear consistently in all ON pointings and no OFF pointings. 

It has been noted by the BL team \citep{Enriquez:2017:turboSETI_limitations} and others \citep{Margot:2021:} that {\tt turboSETI} has two potential weaknesses. First, frequency binning causes a significant drop in sensitivity for drift rates above 0.16 Hz/s. Fortunately, for these observations the expected drift rates are of the same order as the drift rate where this drop in sensitivity occurs. Thus, we do not expect this issue to cause significant sensitivity loss in this case. Second, it has been suggested that the filtering algorithm within {\tt turboSETI} may be prone to missing signals that overlap with easily identifiable RFI.

However, the narrow range of possible drift rates predicted by our theory allows for a much more sensitive search \citep{Wright2020}. We were able to inspect all of the signals within this range by eye (3864 hits in L-band and 448 hits in S-band) to ensure that no signals of interest were discarded along with RFI during filtering.

\subsection[L-band]{L-band}

\begin{figure}
\centering
\begin{minipage}{0.9\textwidth}
\centering
\includegraphics[width=\textwidth]{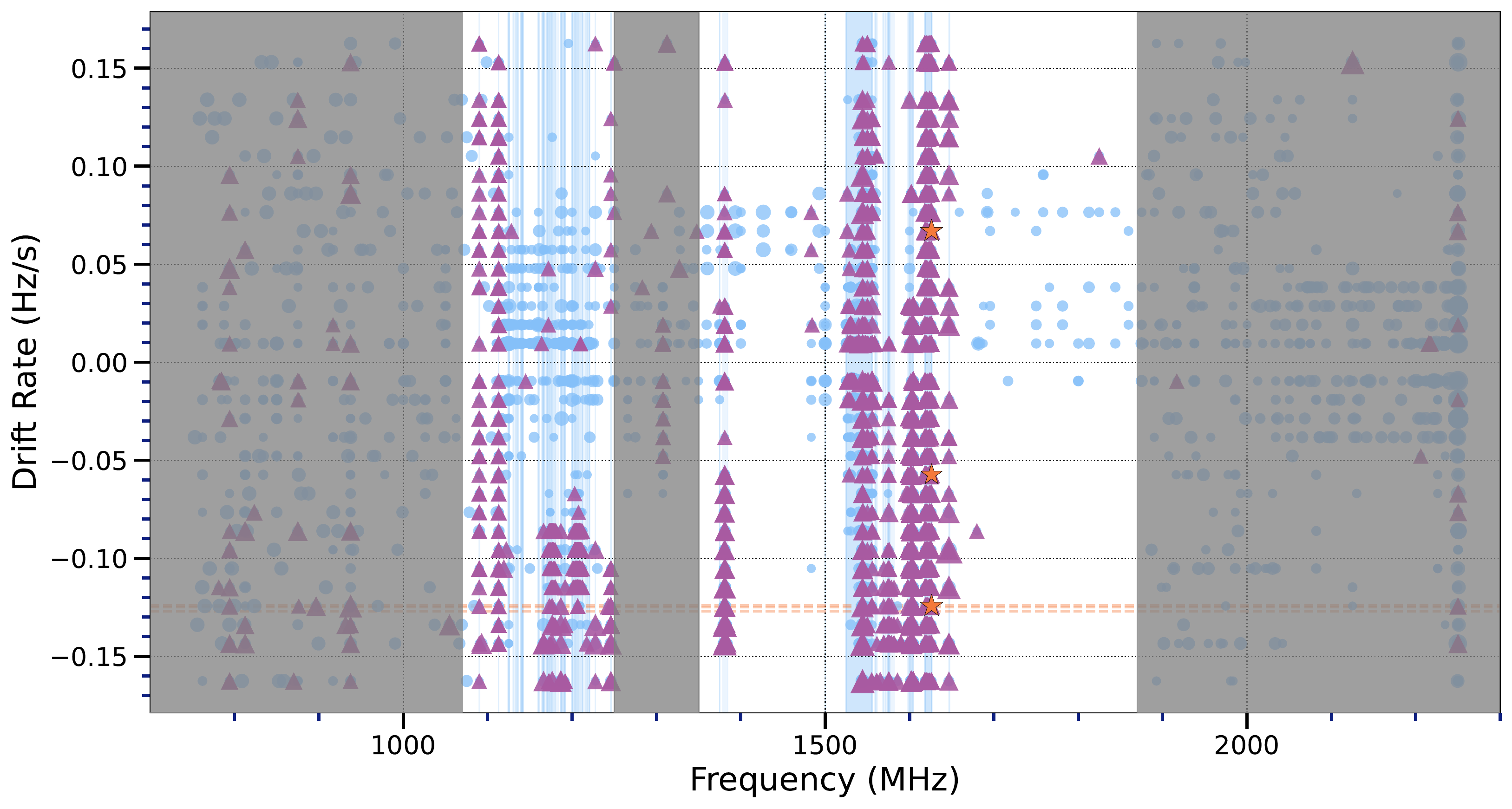}
\caption{All of the signals above a SNR of 10 detected at the first filter threshold (any signal above the SNR detected in any ON source pointing) of {\tt turboSETI} in L-band. Each signal is sized by the log of the SNR. The drift rates of each signal are determined by the {\tt turboSETI} algorithm. Hits passing only the first filter threshold (any signal above the SNR detected in any ON source pointing) are marked as blue dots. The purple triangles indicate hits passing the second filter threshold (any signal above the SNR in at least one ON and no OFF source windows). The golden stars indicate the hits that passed the third filter threshold (signal detection in all ONs and no OFFs). The orange dashed lines correspond to the barycentric drift rates, of -0.124, -0.125, and -0.127 Hz/s, for each of the 3 ON source pointings. The faint blue vertical stripes correspond to signals within 1.87 MHz (0.1\% of the maximum frequency in this band) of each other, likely indicating they are part of a RFI comb. Data outside regions of detector sensitivity as well as those falling within the notch filter (1.25-1.35 GHz) have been greyed out.}
\label{fig:L_freqs}
\end{minipage}
\end{figure}

The band pass filter used on GBT in L-band has sensitivity over the range of about 1.07 to 1.87 GHz, with a notch filter from 1.25 - 1.35 GHz. 
The RFI environment for L-band is notoriously ``noisy," and the notch filter eliminates some of the powerful RFI that frequently dominates the band \footnote{\url{https://www.gb.nrao.edu/~glangsto/rfi/lband/}}. The ``hits" plotted in Figure \ref{fig:L_freqs} show the crowded RFI environment. Hits outside regions of detector sensitivity, including the notch filter, have been included in greyed out regions to provide a more complete picture of the data set and RFI environment. 

The abundance of signals detected with multiple drift rates within narrow ranges of frequencies is characteristic of RFI. We refer to signals with the same morphology spanning multiple frequencies as combs. Although {\tt turboSETI} is designed to not re-trigger on the same signal, it often triggers at multiple drift rates on the same comb when that comb spans a large enough frequency range and its morphology is sufficiently complex. The addition of the light blue vertical lines in Figure \ref{fig:L_freqs} illustrates the presence of these combs by highlighting hits closely grouped in frequency, separated by less than 0.1\% of the maximum frequency in the given filter. The orange dashed lines show the expected drift rates for the ON source pointings as discussed in Section \ref{sec:tseti}.

\begin{figure}
\centering
\begin{minipage}{0.9\textwidth}
\centering
\includegraphics[width=\textwidth]{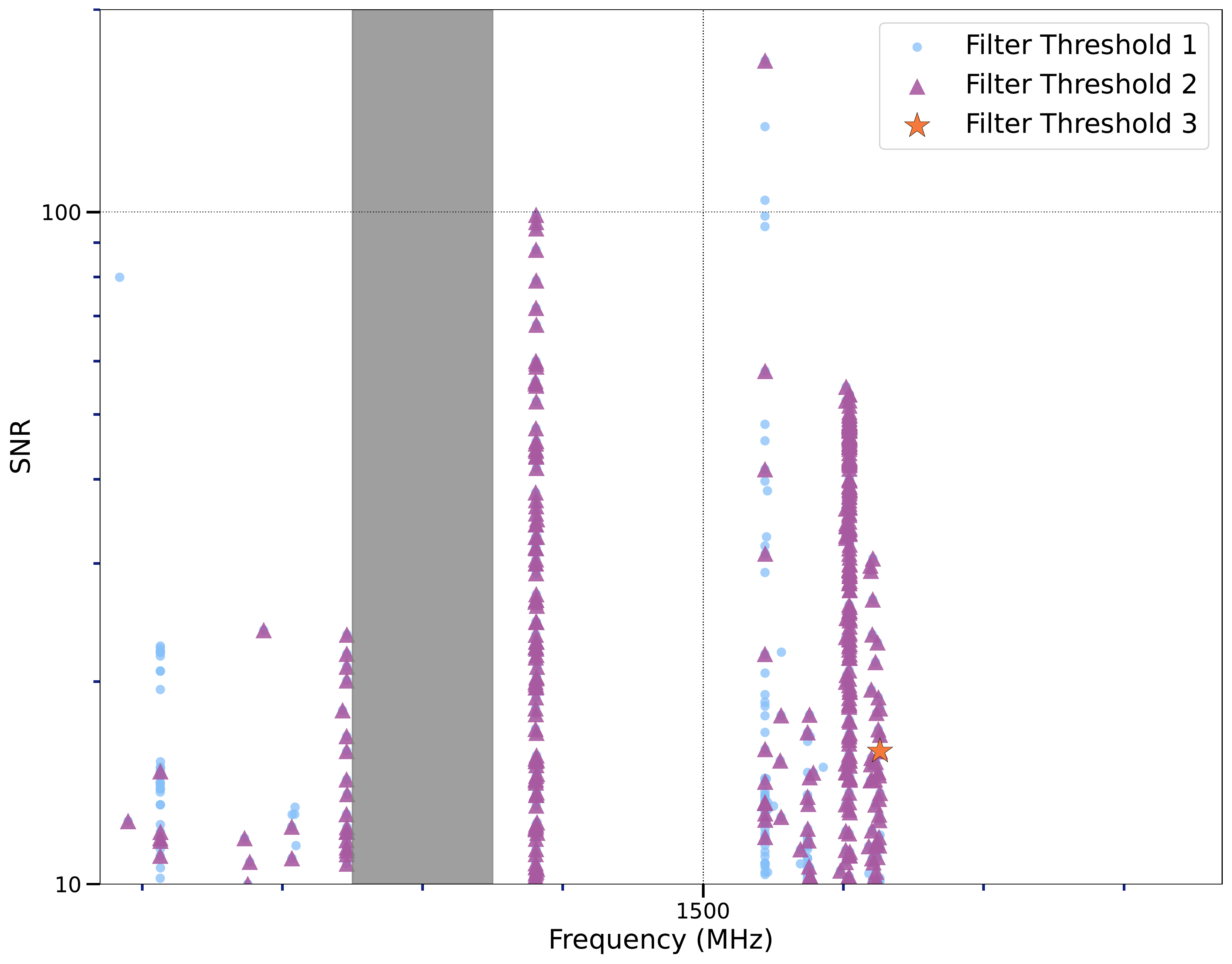}
\caption{All of the signals above a SNR of 10 detected using the {\tt turboSETI} algorithm around the barycentric drift rates, between -0.124 Hz/s and -0.127 Hz/s in the ON target pointings, for these observations in L-band. This shows the SNR for all the hits around the dashed orange lines shown in figure \ref{fig:L_freqs} within the detector sensitivity range. Hits passing only the first filter threshold (any signal above the SNR detected in any ON source pointing) are marked as blue dots. The purple triangles indicate hits passing the second filter threshold (any signal above the SNR in at least one ON and no OFF source windows). The lone golden star indicates the one hit within this drift rate window that passed the third filter threshold (signal detection in all ONs and no OFFs). Figure \ref{fig:L_f3_hit} shows the dynamic spectrum of this signal.}
\label{fig:L_DRange}
\end{minipage}
\end{figure}

From the entire L-band data set, {\tt turboSETI} identified 27995 hits passing the first filter threshold, 23306 of which were within the frequency ranges the band pass filter is sensitive to, as described above and shown in the white regions of figures \ref{fig:L_freqs} \& \ref{fig:L_DRange}. 9929 hits passed the second filter threshold, 9585 of which were within the frequency ranges the band pass filter is sensitive to, and only 3 signals made it through the third filter threshold, all within the sensitivity range.

Figure \ref{fig:L_DRange} isolates all the hits within the barycentric drift rate window shown in Figure \ref{fig:L_freqs}. Only the drift rates for the ON source targets are relevant, so we considered only those drift rates calculated by {\tt turboSETI} near the expected values. The marker types indicate the highest filter threshold that each hit passed. 
Out of the 3 hits passing the third filter threshold, only 1 was within the drift rate range we were searching for. 

\begin{figure}
\centering
\begin{minipage}{0.9\textwidth}
\centering
\includegraphics[width=\textwidth]{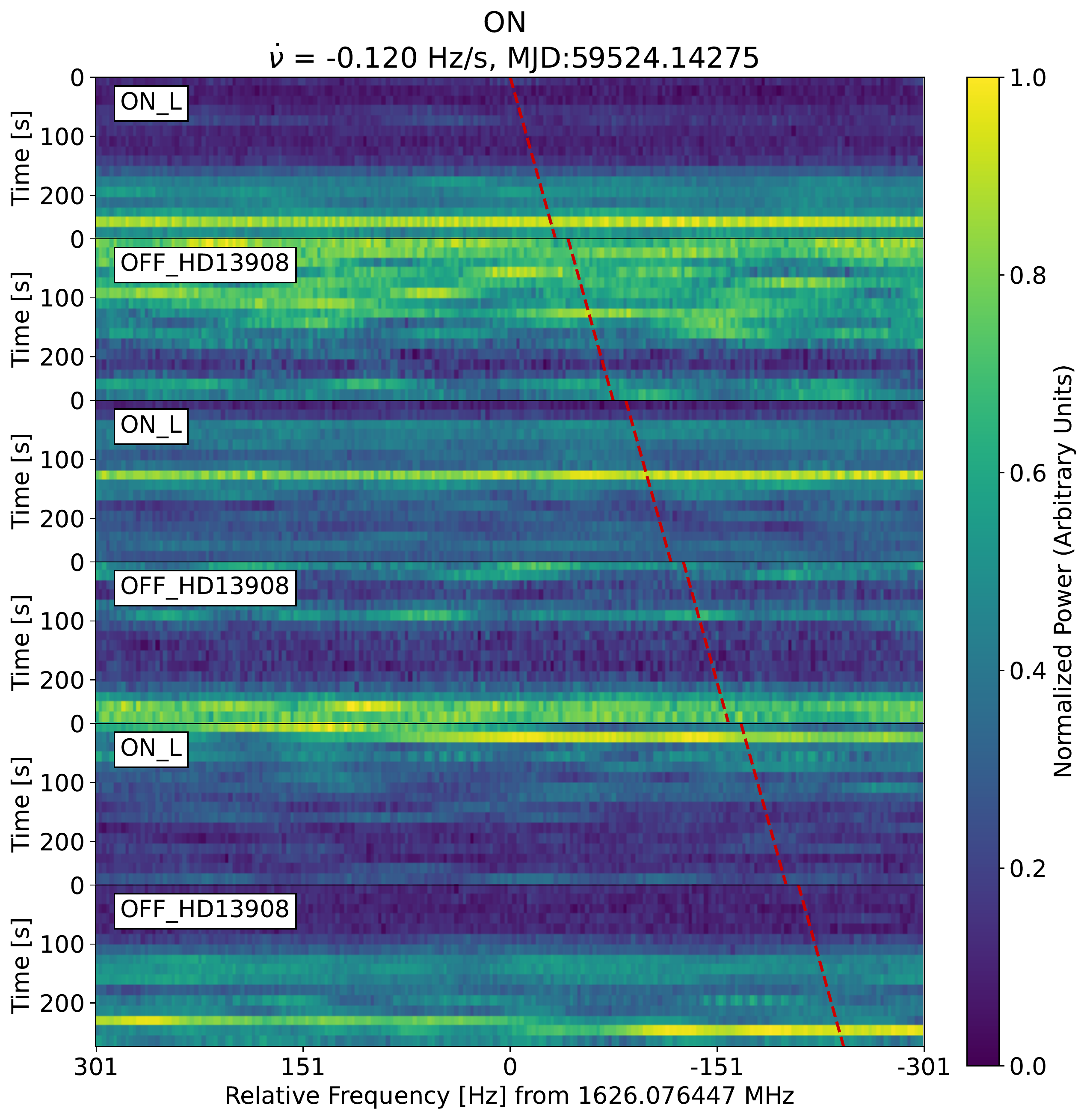}
\caption{The single L-band hit that passed turboSETI’s strictest filtering threshold, intended to select for narrowband signals that appear in every ON source pointing and none of the OFF source pointings, in the target drift rate range near -0.124 Hz/s. Each rectangular block labeled ``ON\_L" and ``OFF\_HD13908" stacked on top of each other shows the ON/OFF observation cadence. This is plotted over a narrow frequency range of 602 Hz centered at the detected signal frequency of 1626.076447 MHz. The red dashed line indicates {\tt turboSETI}’s best-fit drift rate for this signal. The color mapping indicates power, normalized to the maximum power in each panel. It is difficult to determine what {\tt turboSETI} picked up from this plot alone, but Figure \ref{fig:L_f3_zoom_out} shows that it is RFI with a bandwidth of $\sim$80 kHz. Many hits in {\tt turboSETI} show similar features, where a zoomed-out look clarifies that that the signal is drifting RFI.} 
\label{fig:L_f3_hit}
\end{minipage}
\end{figure}

\begin{figure}
\centering
\begin{minipage}{0.9\textwidth}
\centering
\includegraphics[width=\textwidth]{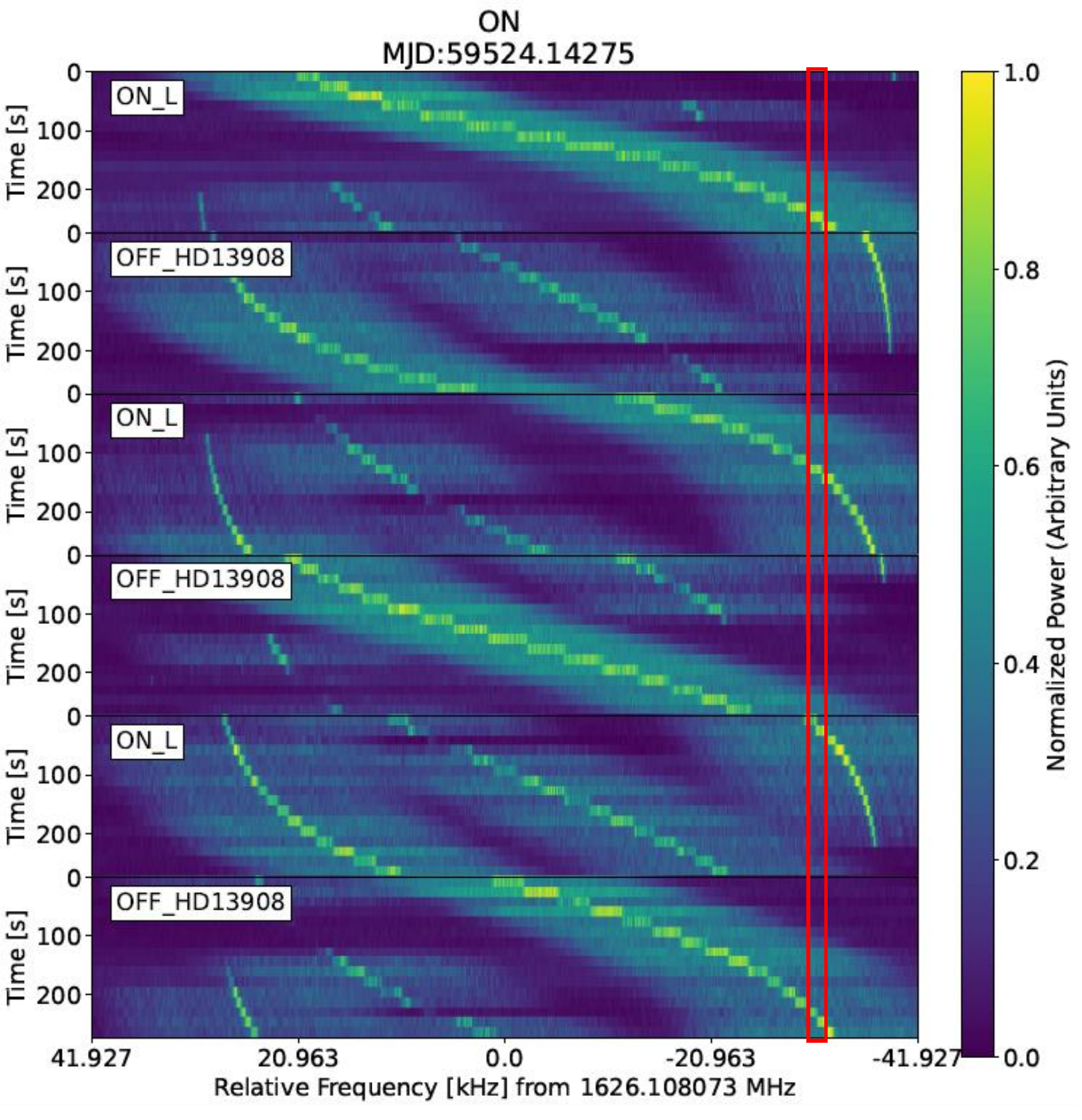}
\caption{Zoomed out plot of Figure \ref{fig:L_f3_hit}, covering more than 80 kHz in L-band (as compared to 602 Hz in the previous plot). The red box indicates the plot window in Figure \ref{fig:L_f3_hit}. From this plot it is easier to see what the algorithm likely picked up as a potential signal. This shows our most promising result is simply an RFI signal.} 
\label{fig:L_f3_zoom_out}
\end{minipage}
\end{figure}

Figure \ref{fig:L_f3_hit} shows the output plot of the dynamical spectrum for this best candidate and more generally illustrates the ON/OFF observing cadence. Figure \ref{fig:L_f3_zoom_out} zooms out to show this signal over a wider frequency range, which clearly identifies both what {\tt turboSETI} likely erroneously triggered on, and that the signal is part of a $\sim$80 kHz wide sweeping RFI signal. While we cannot be certain about the signal's origin, its frequency, brightness, and drift structure are consistent with what would be expected from an IRIDIUM satellite downlinking in its L-band allocation \citep{maine1995}.

To be thorough, we concatenated all of the hits in the data from the ON sources and manually selected them by drift rate, over the range of -0.110 to -0.127 Hz/s. We plotted all 3864 of these signals and examined them by eye to confirm that {\tt turboSETI} only filtered out RFI, finding nothing of interest.

\subsection[S-band]{S-band}

\begin{figure}
\centering
\begin{minipage}{0.9\textwidth}
\centering
\includegraphics[width=\textwidth]{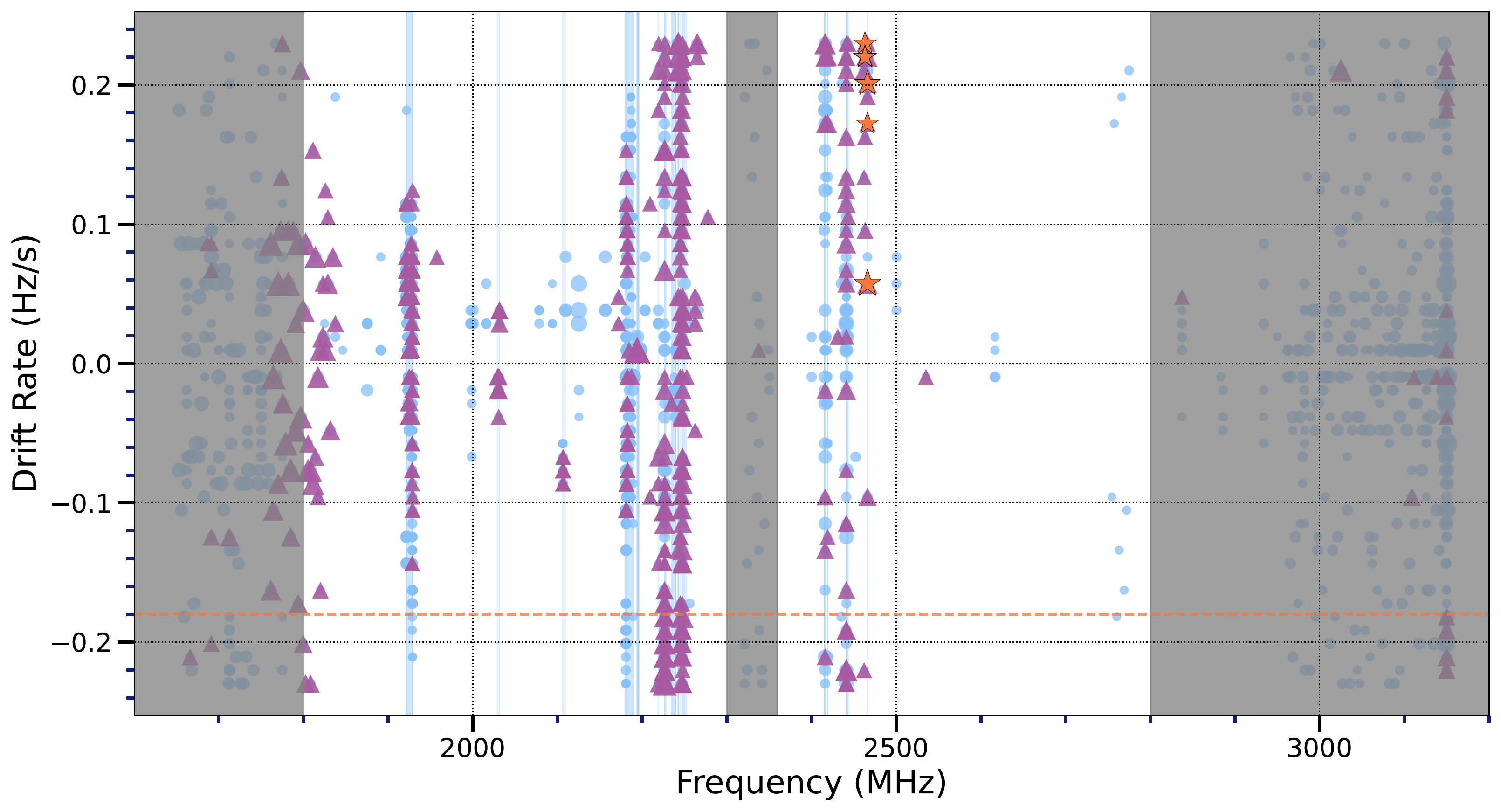}
\caption{All of the signals above a SNR of 10 detected at the first filter threshold (any signal above the SNR detected in any ON source pointing) of {\tt turboSETI} in S-band. Each signal is sized by the log of the SNR. The drift rates of each plotted point, corresponding to a signal above the SNR, are determined by the {\tt turboSETI} algorithm. Hits passing only the first filter threshold (any signal above the SNR detected in any ON source pointing) are marked as blue dots. The purple triangles indicate hits passing the second filter threshold (any signal above the SNR in at least one ON and no OFF source windows). The golden stars indicate the hits that passed the third filter threshold (signal detection in all ONs and no OFFs). The orange dashed line corresponds to the barycentric drift rate, of -0.18 Hz/s, for each of the 3 ON source pointings. The light blue vertical strips correspond to signals within 2.8 MHz (0.1\% of the maximum frequency in this band) of each other, likely indicating they are part of a RFI comb. Data outside regions of detector sensitivity as well as those falling within the notch filter (2.30-2.36 GHz) have been greyed out.}
\label{fig:S_freqs}
\end{minipage}
\end{figure}

The band pass filter used on GBT in S-band is sensitive over the range of 1.80 to 2.80 GHz, with a notch filter from 2.30 to 2.36 GHz \citep{Lebofsky:2019:}. Figure \ref{fig:S_freqs} provides a look at the RFI environment in the S-band observations with all of the hits passing the first filter threshold of {\tt turboSETI} according to their drift rate and frequency. Similar to Figure \ref{fig:L_freqs}, regions where the detector has little to no sensitivity are greyed out. The RFI environment for S-band is not as crowded as L-band, but {\tt turboSETI} still detects multiple frequency combs, again highlighted in vertical bands.

\begin{figure}
\centering
\begin{minipage}{0.9\textwidth}
\centering
\includegraphics[width=\textwidth]{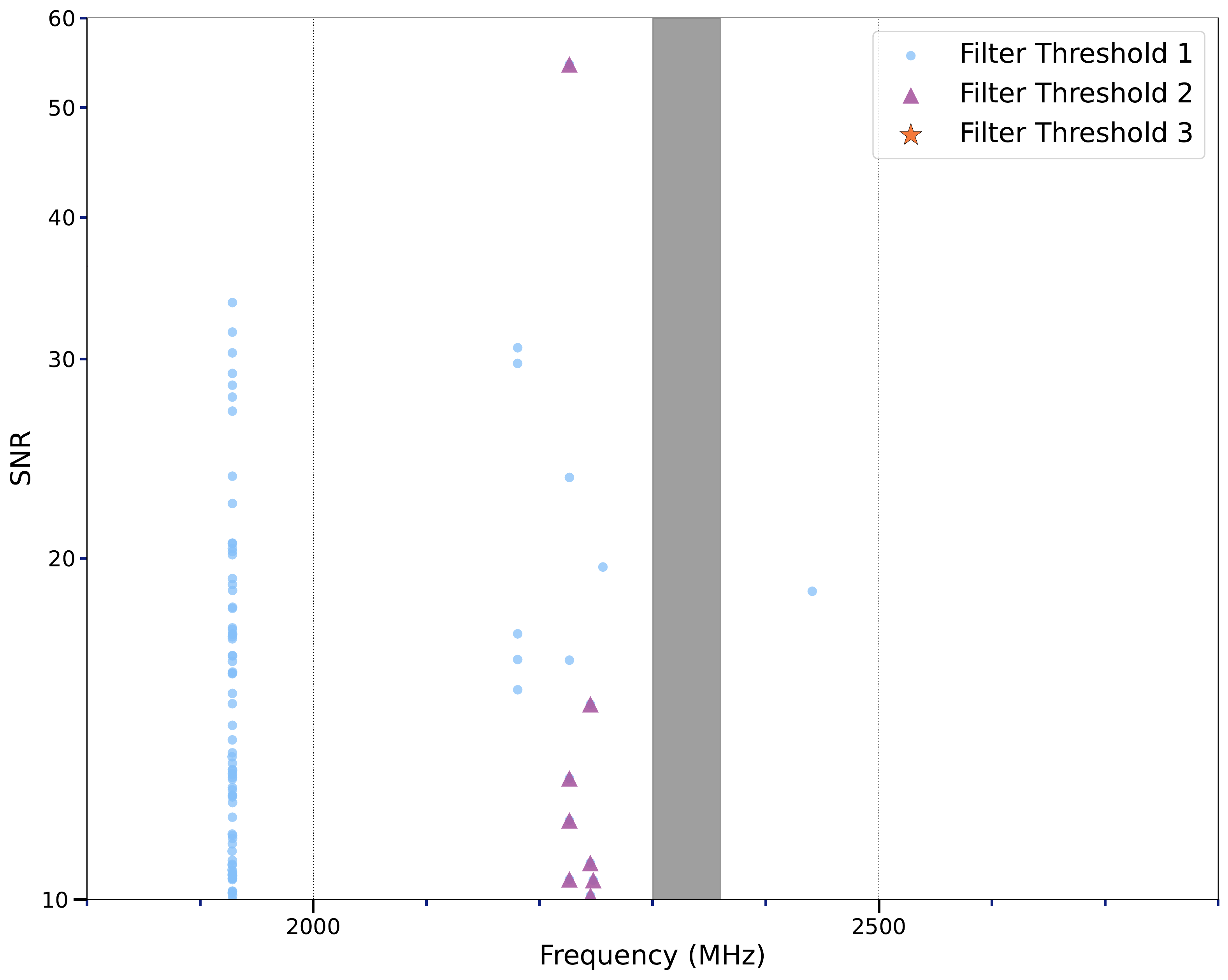}
\caption{All of the signals above a SNR of 10 detected using the {\tt turboSETI} algorithm around the barycentric drift rates, of -0.18 Hz/s, for these observations in S-band. This shows the SNR for all the hits around the dashed orange lines shown in Figure \ref{fig:S_freqs} within the detector sensitivity range. Hits passing only the first filter threshold (any signal above the SNR detected in any ON source pointing) are marked as blue dots. The purple triangles indicate hits passing the second filter threshold (any signal above the SNR in at least one ON and no OFF source windows). No hits within this drift rate window passed the third filter threshold (signal detection in all ONs and no OFFs).}
\label{fig:S_DRange}
\end{minipage}
\end{figure}

Figure \ref{fig:S_DRange} isolates all the hits within the barycentric drift rate window shown in Figure \ref{fig:S_freqs}. Only the drift rates for the ON source targets are relevant, so we considered only those drift rates calculated by {\tt turboSETI} near the expected values. The marker types indicate the highest filter threshold that each hit passed. 
Out of the 6 hits passing the third filter threshold, none were within the drift rate range expected for an SGL probe. 

From the entire S-band data set, {\tt turboSETI} identified 24298 hits passing the first filter threshold, 23563 of which were within the frequency ranges the band pass filter is sensitive to, as described above and shown in the white regions of Figures \ref{fig:S_freqs} \& \ref{fig:S_DRange}. 2226 hits passed the second filter threshold, 2169 of which were within the frequency ranges the band pass filter is sensitive to, and only 6 signals made it through the third filter threshold, all within the sensitivity range.

To be thorough, we concatenated all of the hits in the data from the ON sources and manually selected them by drift rate, over the range of -0.16 to -0.18 Hz/s. We plotted all 448 of these signals and examined them by eye to be sure no signals were being filtered out with the RFI by {\tt turboSETI}. We found no signals of interest within the frequency ranges of this band pass.

\section[Results \& Discussion]{Results \& Discussion} \label{sec:disc}

Our analysis reveals no signals near the expected drift rate for a stellar relay opposite \ACen in L or S bands. Although {\tt turboSETI} output 3 signals in L-band and 6 signals in S-band that passed all of its filters, only 1 of these was found around the expected drift rate, and we determined that all candidate signals are RFI. This was confirmed by visually inspecting the waterfall plots produced by {\tt turboSETI}, paying particular attention to signals around the expected drift rate for each pointing and examining the RFI environment for each band. Figure \ref{fig:L_f3_hit} shows the only one of these candidates identified with a drift rate around what would be expected. Figure \ref{fig:L_f3_zoom_out} shows this to be RFI.

To address concerns that a positive signal could be filtered out with the noise, we examined all of the signals detected around the barycentric drift rate for each band without applying {\tt turboSETI} filter thresholds. Although not practical for many radio SETI searches, our narrow range of drift rates provided the opportunity to scrutinize the data more closely. 
The result of this search provided additional confidence that there were no true positives filtered out by the {\tt turboSETI} algorithm. 

\subsection[HD 13908]{HD 13908}

In addition to our SGL search, we used the OFF source pointings to observe a nearby star, HD 13908. We ran another analysis up to the default drift rate of 10 Hz/s on these observations in L-band and S-band at a SNR of 10. As this was not our primary target of interest, we did not address the limitations of {\tt turboSETI} for this additional target. We report for completeness that no convincing ETI signals were found in these bands using {\tt turboSETI} during the HD 13908 observations.

In L-band, {\tt turboSETI} detected 4932 hits passing the first filter threshold, 1085 passing the second filter threshold, and 7 hits passing the third filter threshold. None of the hits that passed the third filter threshold were of interest.
In S-band, {\tt turboSETI} detected 1733 hits passing the first filter threshold, 392 passing the second filter threshold, and no hits passing the third filter threshold.

\subsection[Detection Sensitivity]{Detection Sensitivity}

Sensitivity for radio SETI observations is often measured with Equivalent Isotropic Radiated Power (EIRP) \citep{Tarter:2001:}, which is a function of the observing instrument, the SNR threshold, the observing time, and distance to the transmitter:

\begin{equation}
    \text{EIRP} = 4 \pi d^2 \times \text{SNR} \times \text{SEFD} \times \sqrt{\frac{\Delta \nu}{2 t}}, \label{eq:EIRP}
\end{equation}
where $d$ is the distance to the transmitter, SNR is the signal to noise ratio, $\Delta \nu$ is the frequency resolution of the detector (where here we assume the transmitter bandwidth is narrower than or equal to the resolution of the detector), and $t$ is the on-source integration time. The System Equivalent Flux Density (SEFD) is specific to the observing instrument in a given band. The SEFD is reported as 10 Jy for the L-band receiver and 12 Jy for S-band at GBT\footnote{\url{https://www.gb.nrao.edu/scienceDocs/GBTog.pdf}}. In both bands, the BL backend achieves a frequency resolution $\Delta \nu$ of 2.79 Hz. The integration time for each of our pointings was 300 seconds, and we analyzed the data with a minimum SNR of 10.

At the shortest theoretical distance of 550 AU, the minimum detectable powers of an isotropic transmitter in L-band and S-band are EIRP$_L$ = 6.6 kW and EIRP$_S$ = 8.1 kW respectively. At the 1000 AU suggested by \citeauthor{Hippke2020b}, EIRP$_L$ = 22 kW and EIRP$_S$ = 27 kW. However, we anticipate a probe utilizing the SGL to be a directed transmitter. Assuming a conservative gain of $10^3$ in L-band and roughly 4 times that in S-band (see Section \ref{sec:Gain}), a $\sim$1 meter transmitter would be detectable by our search above 7 W at 550 AU or 23 W at 1000 AU in L-band, and above 2 W at 550 AU or 7 W at 1000 AU in S-band. These transmitter powers are comparable to cell phones, which emit at 0.6 or 3 W, as well as CB radios at 4 W. A micro-broadcasting FM transmitter puts out 7 W, and Voyager I 23 W.

Using GBT to make observations affords studies of this kind some of the most sensitive radio reception in the world, allowing us to search for relatively low power transmitters. Conducting searches for more powerful transmitters may feasibly be carried out by less sensitive facilities down to the scale of amateur hobbyists.

\subsection[Future Applications]{Future Applications}

Although \ACen belongs to our nearest neighboring star system, it may not be the best star to study in the SGL network context. \cite{Kerby2021} found that close binary star systems are unsuitable for stellar gravitational lens relays due to the high delta-v cost required to maintain alignment. \citeauthor{Kerby2021} instead suggest that the best candidate stars for hosting relay systems are without close companions, without large gas giant planets, relatively low in mass, and not spinning rapidly enough to deviate from sphericity. However, a significant number of stars do not fit this criteria, and the difficulties may often be overcome with some extra delta-v or settling for lesser but still significant stellar lens gains. \cite{Hippke2021} created a first attempt at a prioritized list of targets for these searches, based on factors including the lack of massive companions and the presence of confirmed planets.

Another consideration in selecting stellar relay probe search targets is whether our observations would be made from within the probe's beam, though we have shown that for relatively small spacecraft in radio wavelengths this is not of much concern. Focusing on stars in Earth's transit zone \citep{Heller2016, Sheikh2020, Kaltenegger2021} would allow for observations guaranteed to be in such a beam when Earth is eclipsed by the Sun from the star's viewpoint, even for large optical laser transmitters.

We observed the SGL position opposite \ACen over a single night. Such a limited search may miss extant probes if they are not constantly transmitting. A thorough search for stellar relay probes in the Solar System would involve the monitoring of the antipodes of many stars over longer periods of time. Some progress on this may, in fact, be possible with serendipitous archival observations that happen to include the antipodes of nearby stars (see Palumbo et al., in prep). 

Our search was limited to the region of sky that would be occupied by probes at or past the Sun's focal point of 550 AU. However, \cite{Gillon2021} propose that searching for closer probes may still be worthwhile. \citeauthor{Gillon2021} propose that an array of 1 m lasers may be placed much nearer the Sun, because even at 10 AU, the loss in gain is only a factor of 33. Such transmitters would be subject to stronger perturbations due to planets, but they would also receive much more solar irradiation for power.

As discussed in Section \ref{sec:whatlookfor}, wavelengths longer than 100 $\mu$m are subject to signal strength loss due to the solar corona. \cite{Hippke2020b} suggests that the optimal wavelength range for SGL signals would be between 100 $\mu$m and 1 nm. We observed at relatively long radio wavelengths, 1-4 GHz or 30-7.5 cm. Here, losses due to the solar corona are not prohibitive, but are nevertheless significant. 

The difficulty with using {\tt turboSETI} to analyze observations at shorter wavelengths is, as mentioned in Section \ref{sec:tseti}, that the 1-to-1 binning ratio is 0.16 Hz/s. The drift rate of a signal is a function of the frequency of that signal, so higher frequencies (shorter wavelengths) increase drift rate beyond the 0.16 Hz/s limit, which causes a significant reduction in sensitivity \citep{Margot:2021:}. Sensitivity loss is a trade-off that may be worthwhile for the opportunity to cover more parameter space, especially when specific power levels are not well motivated. It is also possible to recover some of the lost sensitivity for higher drift rates through ``frequency scrunching," wherein frequency bins are effectively summed together. The next generation dedoppler and hit search algorithm from the BL team is in production with the intention to improve search sensitivity.

\section[Summary \& Conclusion]{Summary \& Conclusion} \label{sec:concl}

In this work, we have have described the history and physics behind the proposed use of stars as gravitational lenses to create an interstellar communication network, which motivates an artifact SETI search for such probes in our own solar system. We presented our method for searching for radio transmission from a probe in the solar system in communication with nearby stars and demonstrated this method on \ACene. 
Our analysis found no signals near the expected drift rate during our observations. Therefore, if such a probe exists, it is not transmitting constantly in the 1-3 GHz range. We also present statistics regarding the radio frequency interference environment in this region of sky in the L and S bands. In the future, both archival data and new observations can use similar methods to further search for relays that make use of the Sun's lensing ability for interstellar communication.

\acknowledgements 

\noindent N.T.\ and M.J.H.\ are co-first authors of this manuscript. This paper is a result of the class project for the 2020 graduate course in SETI at Penn State. 
We thank the members of the 2021 undergraduate Penn State SETI course for participating in these observations, and the support of the Green Bank Observatory staff for supporting these classes' visit. J.T.W.\ acknowledges useful conversations with Michael Hippke on the physics of the SGL.

N.T.'s contribution to this material is based upon work supported by the National Science Foundation Graduate Research Fellowship under Grant No. DGE1255832. M.L.P.\ acknowledges the support of the Penn State Academic Computing Fellowship. J.L.\ acknowledges the support of the NASA Astrobiology NfoLD grant \#80NSSC18K1140 and the NASA Pennsylvania Space Grant Consortium (\#80NSSC20M0097 and \#NNX15AK06H). Disclaimer: the findings and conclusions do not necessarily reflect the views of NASA. S.Z.S.\ acknowledges that this material is based upon work supported by the National Science Foundation MPS-Ascend Postdoctoral Research Fellowship under Grant No. 2138147.

The Penn State Extraterrestrial Intelligence Center and the Penn State Center for Exoplanets and Habitable Worlds are supported by the Pennsylvania State University and the Eberly College of Science. Computations for this research were performed on Penn State's Institute for Computational and Data Sciences’ Roar supercomputer. 

Breakthrough Listen is managed by the Breakthrough Initiatives, sponsored by the Breakthrough Prize Foundation. This material is based upon work supported by the Green Bank Observatory which is a major facility funded by the National Science Foundation operated by Associated Universities, Inc.


\software{{\tt barycorrpy} \citep{kanodiaBarycorrpyBarycentricVelocity2018}, {\tt turboSETI} \citep{enriquezTurboSETIPythonbasedSETI2019}, {\tt Skyfield} \citep{Rhodes2019}, {\tt Astropy} \citep{AstropyCollaboration2013, AstropyCollaboration2018}, Matplotlib \citep{matplotlib}, NumPy \citep{numpy}}

\bibliography{refs}{}
\bibliographystyle{aasjournal}

\end{document}